# A Socio-geographic Perspective on Human Activities in Social Media

Ding Ma, Mats Sandberg, and Bin Jiang

Faculty of Engineering and Sustainable Development
University of Gävle, SE-801 76 Gävle, Sweden
Email: ding.ma|mats.sandberg|bin.jiang@hig.se

*(Draft: May 2015, Revision: June 2015, March, June, November 2016)*

**Abstract**
Location-based social media make it possible to understand social and geographic aspects of human activities. However, previous studies have mostly examined these two aspects separately without looking at how they are linked. The study aims to connect two aspects by investigating whether there is any correlation between social connections and users' check-in locations from a socio-geographic perspective. We constructed three types of networks: a people–people network, a location–location network, and a city–city network from former location-based social media Brightkite and Gowalla in the U.S., based on users' check-in locations and their friendships. We adopted some complexity science methods such as power-law detection and head/tail breaks classification method for analysis and visualization. Head/tail breaks recursively partitions data into a few large things in the head and many small things in the tail. By analyzing check-in locations, we found that users' check-in patterns are heterogeneous at both the individual and collective levels. We also discovered that users' first or most frequent check-in locations can be the representatives of users' spatial information. The constructed networks based on these locations are very heterogeneous, as indicated by the high ht-index. Most importantly, the node degree of the networks correlates highly with the population at locations (mostly with R-square being 0.7) or cities (above 0.9). This correlation indicates that the geographic distributions of the social media users relate highly to their online social connections.

## 1. Introduction
The rapid advances in information and communication technology, mobile technology and location-aware technology have tremendously reshaped the ways how social, political, economic and transportation systems work in today's globally connected world. The advanced technologies make the human activities significantly transit from the physical space to virtual space, or a hybrid of physical and virtual spaces (Shaw and Yu 2009, Yu and Shaw 2008). The change occurred in large part because virtual space connects people and enables information to flow without spatio-temporal constraints. The digital ties among people give us great potential to further study human activities since social and physical spaces are inherently constituted with one another (Jones et al. 1997). Social network analysis has conventionally been used for geography about human activities at the aggregated level, e.g. in studies about migration and politics (Murdoch and Marsden 1995, Guo 2009). In recent years, the widespread use of social media like Twitter, Gowalla and Brightkite has made it possible to exploit social connections and activities of users using the location-related information (e.g. Scellato et al. 2011, Sui and Goodchild 2011, Hawelka et al. 2014, Li et al. 2014). Location-related information has been developed into diverse forms, ranging from check-in locations to various location-embedded media (e.g. video, photo, and text) and has gradually become a useful tool for people to communicate each other. Location-based social media data, therefore, can potentially bridge the gap between social science and geospatial information science (Zheng 2011, Yin and Shaw 2015).

Human activities in location-based social media can be largely represented by online friendship and check-in locations. Many studies have discussed these two aspects separately. On the one hand, social scientists have extensively focused on the topological properties of social connections (e.g. Ball 2012, Newman 2010, and Helbing 2007). On the other hand, geographic researchers have used check-in locations for studying human movement patterns. For example, previous studies (e.g., Takhteyev et al. 2012, Onnela et al. 2011, Kulshrestha et al. 2012) took both social and geographic aspects into account by leveraging location information to determine whether geographic distance influenced social connections. Most recent studies on natural cities derived from massive check-in locations (Jiang and Miao 2015, Jiang 2015a) have demonstrated strikingly scaling structures and nonlinear dynamics of human activities. However, few studies investigated check-in locations through social connections, i.e., how check-in locations associates with social connections. While some studies have indeed looked into the social ties under the check-in locations (Cho et al. 2011, Scellato et al. 2011), they concentrated on how social connections affect human mobility pattens, rather than the mutual relationship between social and geographic aspects. This paper aims to link the social connections and check-in locations together from a socio-geographic perspective.

Social connections and check-in locations are accumulated and further developed based on tens of thousands of people of various socio-economic backgrounds. The related information grows very fast and is recorded at very fine spatiotemporal scales. The study therefore can be situated in a big data context. Big data is usually characterized by high volume, high variety, and high velocity (Mayer-Schonberger and Cukier 2013) and it has great impacts on social science in terms of how we conduct related research (Lazer et al. 2009). Big data differs from conventional small data that is usually collected and maintained by statistical and census authorities. For example, big data is measured and collected at an individual level, while conventional small data is usually estimated and aggregated; big data means the entire data set so called population, whereas small data mostly refers to a sample of the population (Jiang and Thill 2015). Based on these characteristics, big data should be considered to be a new paradigm both statistically and geometrically (Jiang and Thill 2015, Jiang 2015b). Coming from diverse individuals, big data is likely to be very heterogeneous. The heterogeneity can be characterized by a power-law distribution. Power-law distribution is the most typical heavy-tailed distribution. If data is heavy tailed, the scaling hierarchy can be uncovered and visualized by head/tail breaks (Jiang 2013, Jiang 2015). Head/tail breaks derives the inherent hierarchies by recursively splitting up the data around the average as long as they are heavy-tailed distributed. The number of recurring times of the scaling pattern that small things are a majority, while large things are a minority is defined as the ht-index (Jiang and Yin 2014), which can measure the heterogeneity of the big data.

This study aims to explore whether there is any correlation between social connections and check-in locations from locaton-based social media data. Relying on some complexity science methods (see Section 3), we examined the heterogeneity of check-in locations and user-location relationships. We found that user check-in patterns are very heterogeneous at both individual and collective levels. This heterogeinety is measured by power law distributions and ht-index. Next, we built up natural cities based on the users' first or most frequent check-ins locations and discovered that they are the most representative locations and reflect more than 80% of total check-in locations. Finally, the location-location/city-city connections were subsequently established through user social connections (see details in section 3.2). We subsequently found that users' social connections relate highly to their spatial distribution through the correlation between the location/city metrics (e.g. the population at each check-in location or city) and its degree of socio-geographic networks.

The remainder of this paper is organized as follows. Section 2 introduces two datasets from Brightkite and Gowalla regarding their sizes and characteristics. Section 3 briefly describes some complexity science methods used for data analysis, and how the various networks were constructed. Section 4 presents the scaling characteristics of locations and user check-in patterns, and shows both analytic and visualization results of constructed networks. Section 5 further discusses the implications of this study. Finally, section 6 concludes and points to future work.

## 2. Data

The data come from the location-based social media Brightkite and Gowalla, both of which started in 2007 and ended in 2012. These social media were similar to the most popular location-based social medium Foursquare but without gaming features. They were primarily intended for networking registered users all over the world via their visited places; namely, check-in locations from mobile devices. Users could establish mutual friendship connections, share locations and photos, and leave comments each other. The location was a useful tool through which users check other nearby users and see who had been there before. In Gowalla, users could also check a user's recent history in a given place (Scellato et al. 2011). Both social media provide public APIs to obtain user's friend list and check-in records.

Table 1: Basic statistics on data sets from Brightkite and Gowalla in the US
(Note: The number in the bracket indicates the total numbers in the world, from which the US occupied a majority.)

|  | Brightkite | Gowalla |
| --- | --- | --- |
| User | 30,927 (58,228) | 52,249 (196,591) |
| Social connection | 113,770 | 222,016 |
| Check-in Location | 2,780,042 (4,491,143) | 3,303,981 (6,442,890) |
| Unique Location | 404,174 (772,968) | 599,846 (1,280,969) |
| Timespan(Month) | 31 | 20 |

The dataset of each social media contains two parts: social connections and user check-in records (data available at http://snap.stanford.edu/data/index.html#locnet). We chose U.S. mainland area for our study because it was the most popular area in terms of both the number of users and the number of check-ins. Social ties and check-in data from Brightkite and Gowalla were global in their coverage, so we extracted users' social network and check-in locations in U.S.. As shown in Table 1, the U.S. constitutes the largest graph, consisting of tens of thousands of users and their undirected connections. The check-in records from both social media are in the form of a list of check-in locations containing the user ID, timestamp, XY coordinates, and location ID. Both datasets contain a similar number of check-in locations, but with different timespans. For Brightkite, the data was from April 2008 to October 2010, while for Gowalla, it was a period of one year and eight months from February 2009 to October 2010. The locations of both social media are multiple checked or repeated; for example, 404,174 of the 2.8 million Brightkite locations are unique.

## 3. Complexity science methods

By considering both geographic and social aspects of human activities in social media, this study can be situated in the big data context for analyzing human activities regarding the scaling property and complex networks (Figure 1). We investigated the scaling property of locations and user check-in patterns and constructed the socio-geographic networks by developing social connections into users' first/most frequent check-in locations. We adopted two complexity science methods for data analysis: power law detection based on maximum likelihood (Clauset et al. 2009) and head/tail breaks (Jiang 2013, Jiang 2015a) for classification and visualization. The power-law detection is used for mathematically examining the complexity from the fitness of data with power law distribution, and head/tail breaks for revealing the inherent scaling hierarchies or patterns.

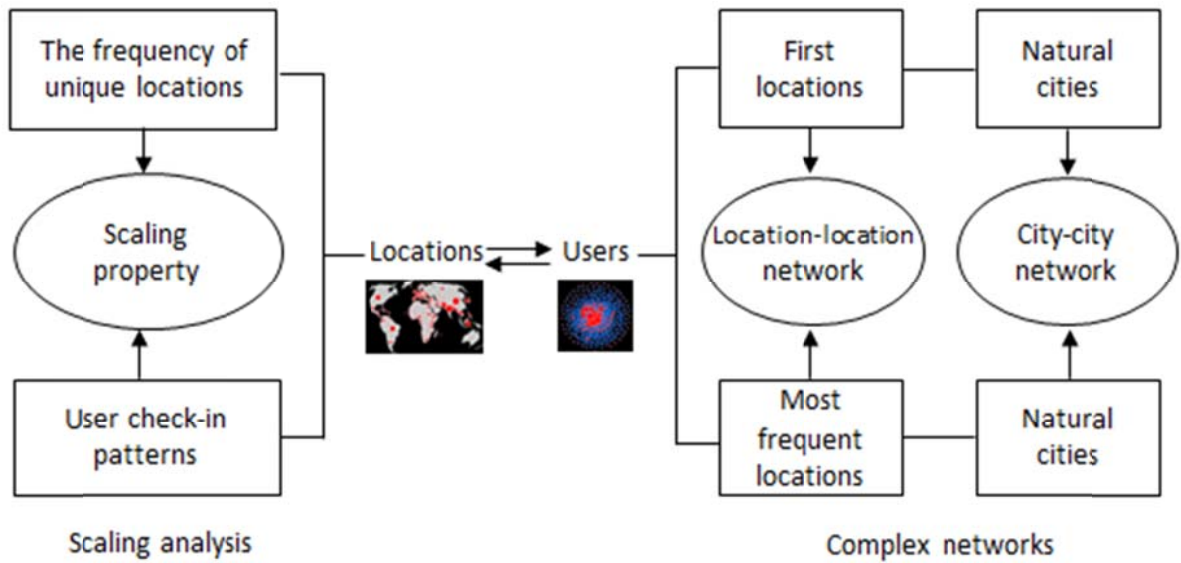

Figure 1: (Color online) The overall struture of this study

**3.1 Scaling analysis on human activities in social media**

A complex system consists of many mutually interacting components, which lead to collective behavior that cannot be simply deducted from the sum of individuals (Newman 2011). A complex system differs fundamentally from a simple system that simple and predictable. A good example of complex systems is the human brain. In this study, the human activities in social media are conducted both socially and geographically, which is reflected by social connections and check-in locations, respectively. Human activities therefore should be analyzed from the perspective of the complexity science. In fact, the check-in activities and networking friends can be considered to be unpredictable and self-organizing processes because they evolve in a bottom-up manner that lacks any central control and is based on the individuals. Therefore, we can better understand the complex structure and nonlinear dynamics of the geographic and social aspects of human activities by adopting the complex science perspective.

The main characteristic of a complex system is its heterogeneity, which can be well characterized or represented by the scaling property (Barabási and Albert 1999, Mandelbrot and Hudson 2004). Scaling or scale-free refers to the fact that there are far more small things than large ones. Data that bears scaling property exhibits a power-law distribution or heavy-tail distribution in general. In the absence of a well-defined mean or scale, it is difficult to analyze such data using traditional methods, such as Gaussian statistics and Euclidean geometry. Instead, complexity science methods such as power law statistics, scaling hierarchy, and complex networks are more appropriate for better understanding the underlying scaling property and its hierarchy. We therefore adopted two methods for scaling analysis in this study: power law detection suggested by Clauset et al. (2009), and head/tail breaks (Jiang 2013). These two methods complemented each other and applied to well characterize and visualize the heterogeneity of geospatial big data (Ma et al. 2015).

A power law distribution indicates the probabilities of a value (y) being proportional to some power of a quantity (x), which is denoted as follows:

$$y = kx^{-\alpha} \qquad [1]$$

where k is a constant and $\alpha$ is the power law exponent.

The well-known examples of power laws are word occurrences in a large document, city sizes, and wealth distributions in a large country, which all follow so called Zipf's law (1949) or Pareto distribution. The simplest way to examine the power law distribution is to take logarithm scale on both axes; that is, $\log(y) = -\alpha \log(x) + \log(k)$. This way a straight line will be obtained if data is power-law distributed. However, this method cannot effectively deal with the disordered tail at the very end of the distribution since the method is essentially based on the ordinary least square estimation. Thus, researchers have developed more reliable statistical tests using maximum likelihood, which has been shown to be more robust for detecting the power-law distributions (Marta et al 2008, Clauset et al. 2009). This method is able to effectively estimate not only the power law exponent ($\alpha$), but also the goodness of fit ($p$). The valid estimated exponent $\alpha$ is between 1 and 3, while the goodness of fit index $p$ is between 0 and 1. Usually, the higher the p value, the better fit. The acceptable threshold for power law fit is commonly set at 0.05.

Head/tail breaks was initially developed as a classification method for data with a heavy-tailed distribution (Jiang 2013). The new classification method splits a data into the head (the small percentage of values above the mean, for example, < 40%) and the tail (the large percentage of values below the mean), and the splitting process goes recursively for the head, until the notion of far more small values than large ones is violated for the head. The derived hierarchy can also measure the complexity of data; that is, the number of times that the scaling pattern of far more small things than large ones recurs. The number of recurring times of scaling pattern plus one is the ht-index (Jiang and Yin 2014). Formally, ht-index (h) can be denoted as:

$$h = m + 1 \qquad [2]$$

where m is the number of the recurring times of far more small values than large ones during the head/tail breaks processes.

Ht-index indicates classes or the hierarchical levels of the data with great heterogeneity. Ht-index is able to measure or quantify the heterogeneity of data: the bigger ht-index value, the more heterogeneous the data is. In this regard, it supplements the power law detection method with the detection of other heavy-tailed distributions (Ma et al. 2015). In other words, if the ht-index value of the data is big enough, the data can be considered as a heavy-tailed distribution even though it may not pass the power law test. Thus head/tail breaks opens up a novel and effective way of analyzing the complexity of big data. Furthermore, for date with a heavy-tailed distribution, the head part is self-similar to the whole data. Therefore, we can recursively take the head for visualization and analysis of the whole data (Jiang 2015a), which can avoid visual clutter while retaining the most core part of the big data.

**3.2 Socio-geographic network construction**
The research built up networks based on user friendship and from the geographic perspective. There are three types of undirected networks in this study: the people–people network, the location–location network and the city–city network. The people–people network is directly from the social connections established in each social medium. A social network with specific layout algorithms can properly visualize such networks; examples include the circular layout (Bertin 1983), and the force-directed layout (Fruchterman and Reingold 1991). For the geographic network, we can assign each user a certain location; that is, the most representative location based on their check-in locations. To determine the user location, we considered the first or most common check-in of each user can be the most representative location of each user.

With the user location and user's social connections, we can further construct the location-location network and the city–city network. In other words, we intend to establish the location-location and city-city connections based on social connections. In a location–location network, each pair of nodes $i$ and $j$ represent two locations which are linked if there are one or more pairs of users who are friends between locations $i$ and $j$. The link is with a weight $w_{ij} \geq 1$ since a location may represent more than one user (some users may have the same first or most frequent check-in), thus the weight equals to the number of pairs of socially connected users. For a city–city network, a city refers to a natural city (Jiang and Miao 2015). The natural city is formed by the clustered check-in location points with short edges (shorter than the arithmetic mean of all edge lengths) under a big triangulated irregular network (TIN) which is composed of all locations in a country. A natural city reflects the basic gathering unit of human settlements or human activities. Each natural city contains a list of check-in locations and users. In such networks, a node represents a natural city, and a weighted edge indicates the number of the pair(s) of friends between two cities.

**4. Results and discussion**

The findings from the analysis applied to the human activities in Brightkite and Gowalla are mostly focused on three aspects. First, user check-in patterns are very complex at both individual and collective level, as captured by the ht-index and power law statistics. Second, the users' first and most frequent locations can be considered as the head part of check-in locations, which reflect to the whole locations. Third, from the constructed socio-geographic networks, social connections correlate highly with users' spatial distribution.

**4.1 Scaling properties of locations and check-in patterns**

We started by inspecting the frequency of unique locations. Both data sets show apparent scaling patterns indicated by the big ht-index values. For example, the ht-index value of Brightkite unique locations is eight which indicates that the scaling pattern occurred seven times. As seen in Table 2, the scaling pattern is very striking because a low head percentage repeatedly occurs almost at each level (below 30%). Next, we examined the scaling property of both data sets in terms of the one-to-many relationship between U.S. users and locations by conducting power law detection. We obtained the number of check-in locations by each user and the number of users at each location. Table 3 presents related power-law fitting metrics such as the exponent and goodness of fit. Both numbers show strikingly power law distributions, which indicates that only a minority of users contributes the majority of check-in locations.

Table 2: Statistics of head/tail breaks on the frequency of Brightkite unique locations
(Node: # = number, Uniloc=Unique location)

| #UniLoc | #head | %head | #tail | %tail | mean |
|---|---|---|---|---|---|
| 404,174 | 45,803 | 11% | 358,371 | 89% | 6.90 |
| 45,803 | 6,975 | 15% | 38,828 | 85% | 46.70 |
| 6,975 | 1,505 | 21% | 5,470 | 79% | 223.85 |
| 1,505 | 358 | 23% | 1,147 | 77% | 693.48 |
| 358 | 92 | 25% | 266 | 75% | 1715.15 |
| 92 | 23 | 25% | 69 | 75% | 3668.67 |
| 23 | 8 | 34% | 15 | 66% | 8027.83 |

Table 3: Power-law fitting statistics of one-to-many relationship between users and locations from Brightkite and Gowalla
(Note: α = power law exponent, p = goodness of fit, min and max refer respectively to the minimum and maximum value for number of users/locations for power law detection)

| | Brightkite | | | | Gowalla | | | |
|---|---|---|---|---|---|---|---|---|
| | α | p | min | max | α | P | min | max |
| #Location per user | 1.82 | 0.13 | 70 | 2,100 | 2.61 | 0.26 | 221 | 2,059 |
| #User at each location | 2.48 | 0.89 | 6 | 1,437 | 2.61 | 0.87 | 11 | 2,157 |

Based on these preliminary analysis results, we further studied the user check-in patterns from both individual and collective perspectives. Since there is a one-to-many relationship between user and location, we first calculated the frequency of each location for every single user. This enabled us to characterize each user's check-in patterns by the ht-index value of the frequency of each location at which the user checked-in. Using Brightkite data set as an example, we found that the check-in pattern of an individual user can be very complex. For instance, there was one active user who had 1,263 checking locations, of which the checking frequencies range from 1 to 60 and the ht-index value of those checking frequencies is 6. More interestingly, the scaling pattern also exists at the collective level. The rank size plot (Figure 2) shows that the distribution of ht-indices of all Brightkite users. Overall the check-in patterns of all users exhibit a heavy-tailed distribution, which indicates that the check-in pattern varies greatly from one user to another.

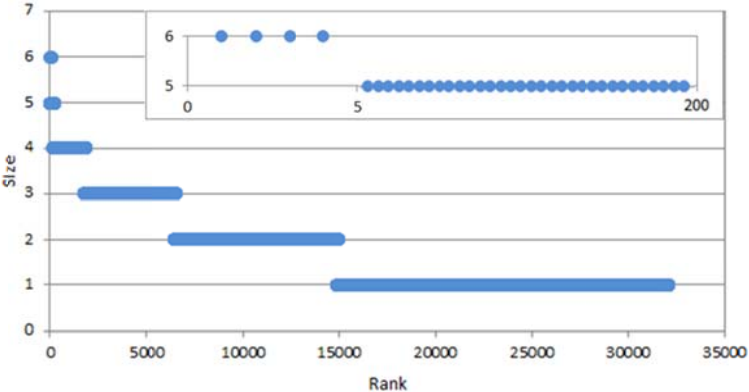

Figure 2: (Color online) The rank-size plot of the ht-index value of location frequencies for each Brightkite user.
(Note: The nested inner plot provides the enlarged view on users with ht-index of 5 and 6.)

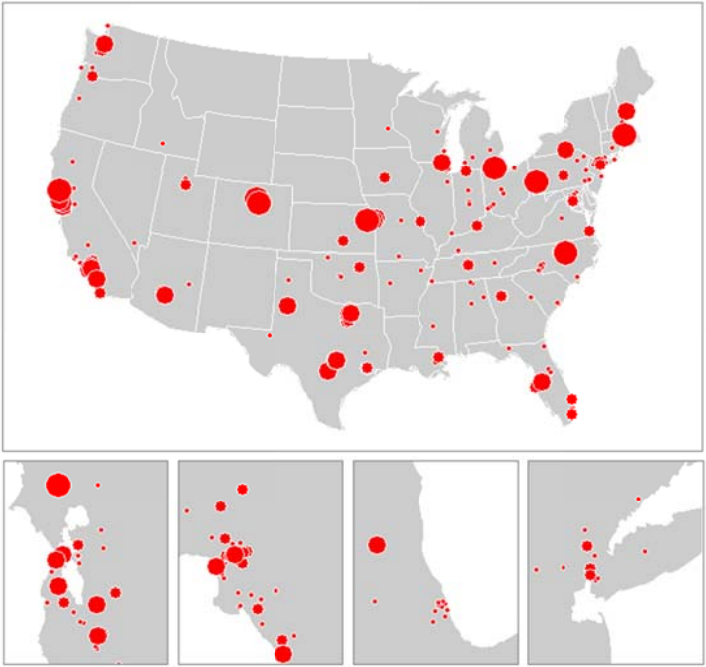

Figure 3: (Color online) Spatial distribution of Brightkite users
(Note: The map shows the top 5 classes including 504 nodes. The dot size is proportional to each user's social degree. The four insets from the left to the right provide the enlarged view from left to right: San Francisco, Los Angeles, Chicago and New York.)

**4.2 First locations, most frequent locations, and natural cities**
Having examined the locations and check-in patterns, we then looked into the first and most frequent locations of each user. Here we selected the Brightkite data set to present the results of our analysis. To begin with, we mapped the first and most frequent check-in location of each user respectively. As a result, these two kinds of locations have an approximately 60% overlap. Figure 3 shows the distribution of users via most check-ins at the top five hierarchies. The top users are mostly from the big city areas such as Los Angeles and San Francisco. The scaling pattern of the users' geographic distribution is akin to that of the city population (Figure 3 in Jiang 2015b). However, some peculiarities exist in that there are barely any users with a large degree in Chicago. Such a pattern also occurs in New York. It could be explained as the lack of market penetration in eastern coast area, for that its biggest contemporary competitor Foursquare was based in New York.

We then built up natural cities from the user locations. From the statistics shown in Table 4, either the first or most common check-in of each user is representative of the whole. Specifically, although both types of user locations are a very small part of unique locations, their resulting natural cities contain more than 80% of all locations. In this regard, they are the head part of entire check-in locations, so that the natural cities based on user locations are also the head of all natural cities. Furthermore, they match more than 80% of the top few classes of natural cities from all check-in locations generated in the previous work (Jiang and Miao 2015), as Figure 4 shows. For further analysis, we calculated the user check-in activity metrics: the population and number of locations for each location and natural city. For each user location, the population is equal to the number of users who checked in there. The number of locations refers to how many times it was checked in. For each natural city, we use the number of users and locations within the city area as its population and its number of locations, respectively.

Table 4: Statistics about natural cities from user locations of Brightkite
(Note: # = number, ULoc = user location, NC = natural city, ULocIn = user locations in the cities, ULocOut = user location outside the cities, Loc = check-in location.)

|  | #ULoc | #NC | #ULocIn | #ULocOut | #LocIn | #LocOut |
|---|---|---|---|---|---|---|
| First check-in | 17,724 | 458 | 15,853 | 1,871 | 2,244,035 | 536,007 |
| Most check-in | 19,450 | 427 | 17,947 | 1,503 | 2,245,979 | 534,063 |

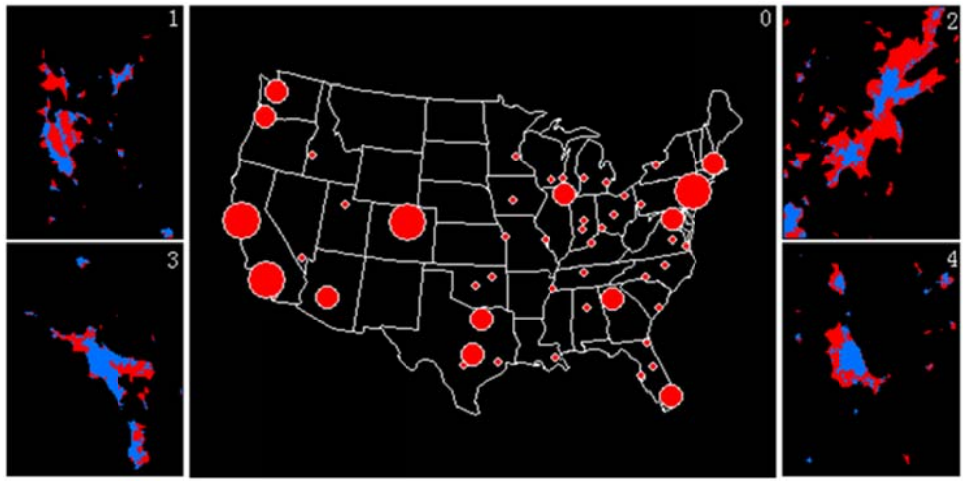

Figure 4: (Color online) The natural cities derived from Brightkite user most check-in locations
(Note: The dot size is proportional to the city population; the red patches represent the natural cities from all check-in locations while the blue ones for cities from user locations. For (a), Panel 0 shows the top four levels of derived natural cities, which totally match the ones from all check-in locations; for example, San Francisco (Panel 1), New York (Panel 2), Los Angles (Panel 3) and Chicago (Panel 4))

### 4.3 Analysis of socio-geographic networks
With the extracted check-ins and natural cities, we then constructed the location–location network and city–city network from social connections, respectively. The people-people network is a scale-free network. Figure 5 shows the network in the force-directed layout as a whole or at each hierarchy regarding the node degree by applying head/tail breaks. Consistent with the statistics of the node degree, the network contains several hierarchies, and each hierarchy possesses the underlying scaling pattern of far more users with a relatively small degree than ones with a large degree. Also, the top classes (even the most top one in panel 4) are self-similar to the whole because they retain the same scaling structure.

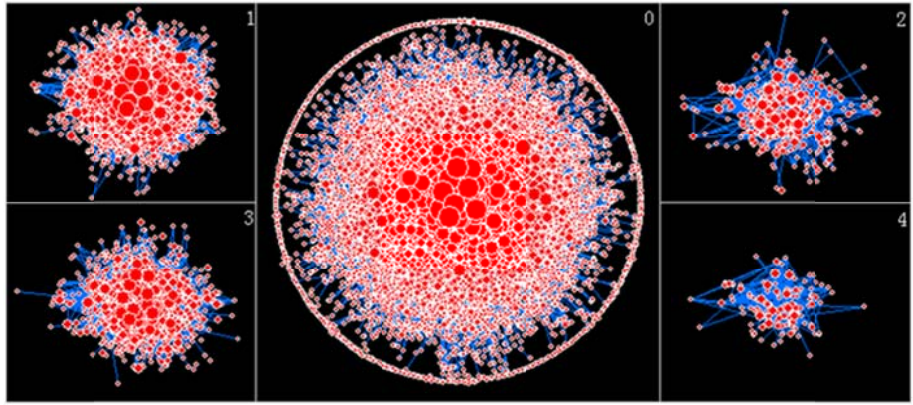

Figure 5: (Color online) Social network of Brightkite data set
(Note: Panel 0 shows the whole Brightkite social network in a force-directed layout, Panels 1–4 show each level of network based on head/tail breaks, the red dots indicate the classified user levels regarding the degree.)

The related metrics of location-location and city-city networks are shown in Table 5. Consequently, both networks possess strongly scaling patterns. For nodes, there are far more small locations/cities than big ones regarding population and location number. For edges, the big ht-index values indicate that there are far more light-weighed edges than heavy-weighted ones for both networks. We foresee that the pattern of both networks containing all classes will certainly be indiscernible in the visualization due to the bigness and complexity of the networks. Therefore, we selected a few top levels regarding edge weights based on head/tail breaks instead of showing the whole network. The constructed networks possess obviously scaling hierarchies as Figure 6 shows. Both user locations and natural cities of top classes locate in big cities such as New York and Los Angeles, and most heavy-weighted edges are among the important locations or big cities (Figures 6a and 6b). More importantly, the weighted degree of the natural city network highly correlates with its population and number of location (with R square being around 0.9), and slightly lower for the location–location network, as Table 5 shows. This result reflects the great correlation between social connections and users' spatial distribution.

Table 5: The related metrics of location–location and city–city networks based on Brightkite user locations
(Note: # = number, ht-index = ht-index value according to edge weight, $R^2$ (population) = the R-square value between the node-weighted degree and the population, $R^2$ (location #) = the R-square value between node-weighted degree and the location number)

|  | First check-in | | Most check-in | |
| --- | --- | --- | --- | --- |
|  | Location–location | City–city | Location–location | City–city |
| # of nodes | 17,724 | 427 | 19,450 | 458 |
| # of edges | 2,787,630 | 8,900 | 3,261,608 | 9,450 |
| ht-index | 9 | 7 | 11 | 7 |
| $R^2$(population) | 0.79 | 0.90 | 0.66 | 0.90 |
| $R^2$(location #) | 0.49 | 0.91 | 0.39 | 0.91 |

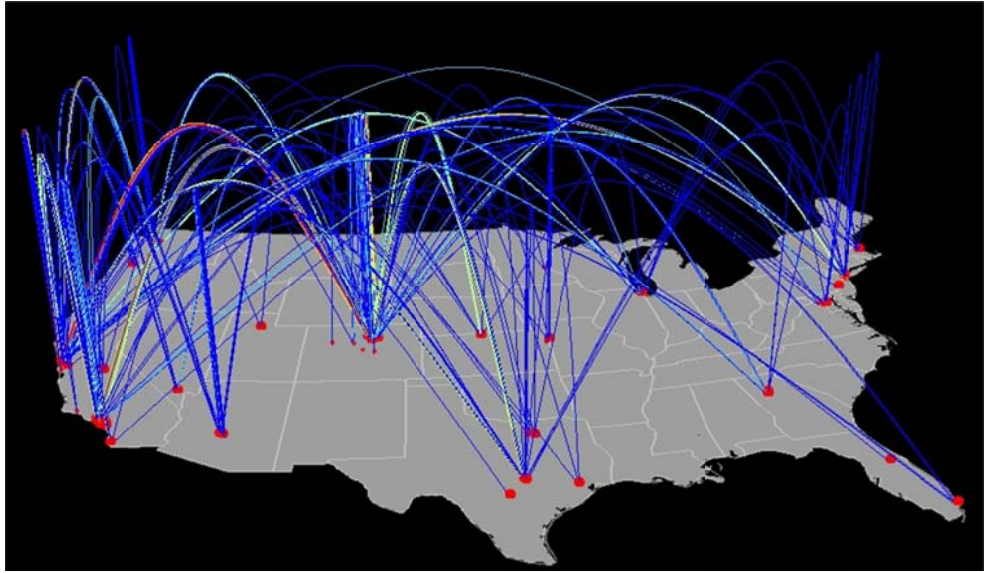

(a)

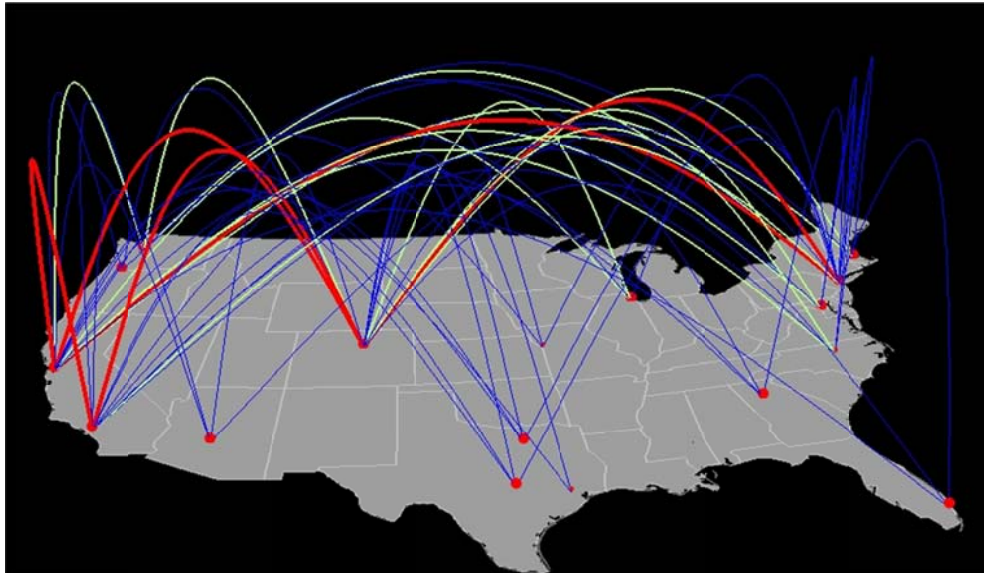

(b)

Figure 6: (Color online) The location–location network (a) and the city–city network (b) derived from Brightkite user most check-in locations (Note: As mentioned in this paper, only the top levels based on head/tail breaks are shown for visual clarity. Edges are visualized by spectral colors according to weights, with blue as the lowest weight and red as the largest weight. The map (a) selects the top four hierarchies and (b) for the top three levels regarding the edge weight.)

We then conducted the same analysis on Gowalla dataset. Gowalla is a more active social medium, for the number of check-in locations in Gowalla is more than that of Brightkite, but the time span is less than almost one year. With more users and check-in locations, the resulting networks are much bigger than the ones from Brightkite, especially the location-location network, which has about ten times as many edges as Brightkite. As Table 6 shows, the node degrees of constructed networks have better correlations with the population and number of locations at both user locations and natural cities, except the location–location network based on users' first check-ins. Therefore, we believe that as the increasing amount of location-based social media data becomes available, both the location–location and city–city networks can better mirror the inter-relationship between people's spatial activity and social connections.

Table 6: The related metrics of the location–location and city–city networks based on Gowalla user locations
(Note: # = number, ht-index = ht-index value according to edge weight, $R^2$ (population) = the R-square value between the node-weighted degree and the population, $R^2$ (location #) = the R-square value between node-weighted degree and the location number)

|  | First check-in | | Most check-in | |
| --- | --- | --- | --- | --- |
|  | Location–location | City–city | Location–location | City–city |
| # of nodes | 41,818 | 923 | 38,788 | 883 |
| #of edges | 45,110,817 | 16,770 | 34,412,469 | 15,080 |
| ht-index | 12 | 7 | 13 | 6 |
| $R^2$(population) | 0.32 | 0.97 | 0.77 | 0.94 |
| $R^2$(location #) | 0.52 | 0.93 | 0.67 | 0.90 |

## 5. Further discussion on this study

The increasing availablity of massive geospatial data helps us to analyze and better understand human activities at both an individual and collective level at the country or even global scale, through the check-in locations. These natural cities, to a great extent, reflect a portrait of human activities. Based on these natural cities, we can identify the fractal structure and nonlinear dynamics of urban systems (Jiang and Miao 2015, Jiang 2015a) and further find the interrelationship between social networks and geographical space. In this section, we add further discussions on this study, related to big data, and geospatial analysis in the era of big data.

This study did not consider possible sampling biases, given the fact that large location-based social media data differ fundamentally from conventional small data. There is little doubt that both the users and check-in locations reflect the activities of certain types of people instead of all population. However, we think that the millions of check-ins and tens of thousands of users can be a very good approximation for studying human activities. It can be seen that the user distributions (Figure 3), the derived natural cities (Figure 4), and the constructed networks (Figure 6) basically reflect the situation in reality. For example, it is hard to negate that the big natural cities do not represent the real ones like New York, and Los Angeles. Additionally, In line with Anderson (2008) that "*with enough data, the numbers speak for themselves*", this study was not to investigate whether the user check-in behavior reflect all human beings on the globe, which may demand a perfect data set, but to present only what the data set tells us, i.e., the scaling patterns of check-in locations and users, and the high correlation between social ties and user locations.

Following some complex system thinking, this study aimed to illustrate the underlying complexity and scaling hierarchy from a socio-geographic perspective. We mainly relied on the head/tail breaks and found that both the data and results possess clearly the scaling property. Examples include the user check-in patterns, edge weights of socio-geographic networks, and the population at user locations and natural cities. In this study, head/tail breaks worked as an effective and efficient visualization and analysis tool to obtain the underlying scaling hierarchies. It recursively filters out the data by keeping the important part (head) and removing the less-important one (tail). Because data bearing the scaling pattern is self-similar, the remaining head data can still reflect the whole. In practice, we managed to measure the complexity of check-in locations and networks, and visualize them with good clarity by selecting the head classes.

The study sought for the correlation between social connections and check-in locations. We linked social connections and the users' first and most common check-ins and found the close relationship between them. In terms of location-location networks, it was interesting to find that in Brightkite, the network based on first check-ins has a higher correlation value than that based on most check-ins, while it is just the opposite in Gowalla (Table 5 and 6). This finding indicates that either the first check-in or most frequent check-in is possible to be a good proxy of the user location. From city-city networks, it can be observed that there are more connections among big cities (Figure 6). This observation indicates that the number of social connections does not correlate well with geographic proximity, but depends on the characteristics of a place. In this respect, the socio-geographic perspective can deepen our insights on human activities in social media.

The head part of check-in locations can accurately characterize the entire locations of location-based social media. This is very much like the relationship between city and country. In a big country, cities represent its key aspects, such as economics and culture, but the total area of these cities is usually relatively small compared to the country size. Previous research (Jiang and Liu 2012) found that the total area of city blocks is, on average, less than 10% of that of a country, but that the number of city blocks accounts for approximately 90% of all blocks. In this regard, city blocks are the head part from which we can spot the main characteristics of a country. Note that analyzing the tail part can also be important. For example, the natural cities are actually collective patches of small TIN edges, which are the tail part. Overall, head/tail breaks method helps us to find the essential information by locating the head and tail parts of the big data effectively, and thus it has promising implications for big data mining and analytics.

## 6. Conclusion

This paper investigated human activities in the former location-based social media platforms Brightkite and Gowalla from a socio-geographic perspective. We extracted each user's first or most check-in locations and then used them to build up natural cities. Based on the social connections and obtained geographic locations, we constructed the socio-geographic networks for both data sets. We found that the R square values are overall very high between the node degree of each type of socio-geographic networks and population/location number at both check-in locations and cities. These findings demonstrated that the social connections highly correlate with the users' spatial information and implied that the attempt of enabling both social aspects and geographic space facilitates a better understanding of human activities.

By situating the study in the context of big data, we adopted some complexity science methods such as the power law detection and head/tail breaks in order to develop new insights into big data or human activities. Although they had been previously used, the methods still helped develop new and interesting insights into social and geographic aspects of human activities. Head/tail breaks and its induced ht-index were recursively applied to showcase their advantages in network visualization. They provide us with a powerful analysis tool to effectively uncover the underlying scaling property of locations, user check-in patterns, and socio-geographic networks. Besides, the hidden information and pattern based on the head part of check-in locations imply that the head/tail breaks thinking can be promising for big data mining and analytics. This research can be extended in the future in several directions, and one of which is how the illustrated patterns may change or evolve from place to place and/or from time to time.

**Acknowledgement**:
Bin Jiang's work is partially supported by special fund of Key Laboratory of Eco Planning & Green Building, Ministry of Education (Tsinghua University), China

**References**
Anderson C. (2008), The end of theory: The data deluge makes the scientific method obsolete, *Wired Magazine*, 16(7), http://www.wired.com/science/discoveries/magazine/16-07/pb_theory (accessed May 15, 2016).
Ball P. (2012), *Why Society Is a Complex Matter: Meeting twenty-first century challenges with a new kind of science*, Springer Science and Business Media: New York.
Barabási A. L. and Albert R. (1999), Emergence of scaling in random networks, *Science*, 286(5439), 509-512.

Bertin J. (1983), *Semiology of Graphics: diagrams, networks, maps*, University of Wisconsin Press: Madison.
Cho E., Myers S. A., and Leskovec J. (2011), Friendship and mobility: user movement in location-based social networks, *In Proceedings of the 17th ACM SIGKDD International Conference on Knowledge Discovery and Data Mining*, San Diego, U.S.A, 1082-1090.
Clauset A., Shalizi C. R., and Newman M. E. J. (2009), Power-law distributions in empirical data, *SIAM Review*, 51, 661-703.
Fruchterman T. M. and Reingold E. M. (1991), Graph drawing by force-directed placement, *Software: Practice and experience*, 21(11), 1129-1164.
Guo D. (2009), Flow mapping and multivariate visualization of large spatial interaction data, *Visualization and Computer Graphics, IEEE Transactions on*, 15(6), 1041-1048.
Hawelka B., Sitko I., Beinat E., Sobolevsky S., Kazakopoulos P., and Ratti C. (2014), Geo-located Twitter as proxy for global mobility patterns, *Cartography and Geographic Information Science*, 41(3), 260-271.
Helbing D. (2007), *Managing Complexity*, Springer: New York.
Jiang B. (2013), Head/tail breaks: A new classification scheme for data with a heavy-tailed distribution, *The Professional Geographer*, 65(3), 482 – 494.
Jiang B. (2015a), Head/tail breaks for visualization of city structure and dynamics, *Cities*, 43, 69-77.
Jiang B. (2015b), Geospatial analysis requires a different way of thinking: The problem of spatial heterogeneity, *GeoJournal*, 80(1), 1-13.
Jiang B. and Liu X. (2012), Scaling of geographic space from the perspective of city and field blocks and using volunteered geographic information, *International Journal of Geographical Information Science*, 26(2), 215-229.
Jiang B. and Miao Y. (2015), The evolution of natural cities from the perspective of location-based social media, *The Professional Geographer*, 67(2), 295 – 306.
Jiang B. and Yin J. (2014), Ht-index for quantifying the fractal or scaling structure of geographic features, *Annals of the Association of American Geographers*, 104(3), 530–541.
Jiang, B. and Thill J. C. (2015), Volunteered Geographic Information: Towards the establishment of a new paradigm, *Computers, Environment and Urban Systems*, 53, 1-3.
Jones J. P., Nast H. J. and Roberts S. M. (1997), *Thresholds in Feminist Geography: Difference, methodology, representation*, Rowman and Littlefield: Lanham.
Kulshrestha J., Kooti F., Nikravesh A. and Gummadi P. K. (2012), Geographic Dissection of the Twitter Network, *The International AAAI Conference on Web and Social Media (ICWSM), the 6th International Conference*, Dublin, Ireland, 202-209.
Lazer, D., Pentland, A., Adamic, L., Aral, S., Barabási, A.-L., Brewer, D., Christakis, N., Contractor, N., Fowler, J., Gutmann, M., Jebara, T., King, G., Macy, M., Roy, D., and Van Alstyne, M. (2009), Computation social science, *Science*, 323, 721-724.
Li L., Goodchild M. F. and Xu B. (2014), Spatial, temporal, and socioeconomic patterns in the use of Twitter and Flickr, *Cartography and Geographic Information Science*, 40(2), 61-77.
Ma D., Sandberg M., and Jiang B. (2015), Characterizing the heterogeneity of the OpenStreetMap data and community, *ISPRS International Journal of Geo-Information*, 4(2), 535-550.
Mandelbrot B. B. and Hudson R. L. (2004), *The (Mis)Behavior of Markets: A fractal view of risk, ruin and reward*, Basic Books: New York.
Marta C. G., Cesar A. H. and Barabási, A.L. (2008), Understanding individual human mobility patterns, *Nature*, 453, 779-782.
Mayer-Schonberger V. and Cukier K. (2013), *Big Data: A revolution that will transform how we live, work, and think*, Eamon Dolan/Houghton Mifflin Harcourt: New York.
Murdoch J. and Marsden T. (1995), The spatialization of politics: local and national actor-spaces in environmental conflict, *Transactions of the Institute of British Geographers*, 20(1995), 368-380.
Newman M. E. J. (2010), *Networks: An introduction*, Oxford University Press: Oxford.
Newman M. E. J. (2011), Complex Systems: A survey, *American Journal of Physics,* 79, 800-810.
Onnela J. P., Arbesman S., González M. C., Barabási A. L. and Christakis N. A. (2011), Geographic constraints on social network groups, *PLoS one*, 6(4), e16939.

Scellato S., Noulas A., Lambiotte R. and Mascolo C. (2011), Socio-Spatial Properties of Online Location-Based Social Networks, *The International AAAI Conference on Web and Social Media (ICWSM), the 5<sup>th</sup> International Conference*, Barcelona, Spain, 329-336.

Shaw S. L. and Yu H. (2009), A GIS-based time-geographic approach of studying individual activities and interactions in a hybrid physical–virtual space, *Journal of Transport Geography*, 17(2), 141-149.

Sui D. and Goodchild M. (2011), The convergence of GIS and social media: challenges for GIScience, *International Journal of Geographical Information Science*, 25(11), 1737-1748.

Takhteyev Y., Gruzd A., and Wellman B. (2012), Geography of Twitter networks, *Social Networks*, 34(1), 73-81.

Yin L. and Shaw S. L. (2015), Exploring space–time paths in physical and social closeness spaces: a space–time GIS approach, *International Journal of Geographical Information Science*, 29(5), 742-761.

Yu H. and Shaw S. L. (2008), Exploring potential human activities in physical and virtual spaces: a spatio-temporal GIS approach, *International Journal of Geographical Information Science*, 22(4), 409-430.

Zheng Y. (2011), Location-based social networks: Users, In *Computing with Spatial Trajectories*, Springer: New York, 243-276.

Zipf G. K. (1949), *Human Behavior and the Principles of Least Effort*, Addison Wesley: Cambridge, MA.